\documentclass[conference]{IEEEtran}
\IEEEoverridecommandlockouts
\usepackage{cite}
\usepackage{amsmath,amssymb,amsfonts}
\usepackage{algorithmic}
\usepackage{graphicx}
\usepackage{textcomp}
\usepackage{xcolor}
\usepackage{booktabs}
\usepackage{siunitx}
\usepackage{url}
\usepackage{subfigure}
\usepackage[font=small]{caption}

\def\BibTeX{{\rm B\kern-.05em{\sc i\kern-.025em b}\kern-.08em
    T\kern-.1667em\lower.7ex\hbox{E}\kern-.125emX}}
\begin{document}

\title{Performance Evaluation of a New Scheduling \\Model Using Congestion Window Reservation}

\author{\IEEEauthorblockN{1\textsuperscript{st} Mahsa Noroozi}
\IEEEauthorblockA{\textit{Institute of Communications Technology} \\
\textit{Leibniz Universität Hannover}\\
Hanover, Germany \\
mahsa.noroozi@ikt.uni-hannover.de}
\and
\IEEEauthorblockN{2\textsuperscript{nd} Flavio Gallistl}
\IEEEauthorblockA{\textit{Institute of Communications Technology} \\
\textit{Leibniz Universität Hannover}\\
Hanover, Germany \\
flavio.gallistl@stud.uni-hannover.de}
\and
\IEEEauthorblockN{3\textsuperscript{rd} Majid Noroozi}
\IEEEauthorblockA{\textit{Department of Mathematical Sciences} \\
\textit{University of Memphis}\\
Memphis, USA \\
mnoroozi@memphis.edu}
}

\maketitle
\thispagestyle{plain} 
\pagestyle{plain}

\begin{abstract}
Multipath QUIC is a transport protocol that allows for the use of multiple network interfaces for a single connection. It thereby offers, on the one hand, the possibility to gather a higher throughput, while, on the other hand, multiple paths can also be used to transmit data redundantly. Selective redundancy combines these two applications and thereby offers the potential to transmit time-critical data. This paper considers scenarios where data with real-time requirements are transmitted redundantly while at the same time, non-critical data should make use of the aggregated throughput. A new model called congestion window reservation is proposed, which enables an immediate transmission of time-critical data. The performance of this method and its combination with selective redundancy is evaluated using emulab with real data. The results show that this technique leads to a smaller end-to-end latency and reliability for periodically generated priority data.
\end{abstract}


\section{Introduction}
Real-time is playing an increasingly important role in applications that are dealing with communication via the Internet. These include gaming, video telephony, industrial applications, and cyber-physical systems~\cite{wolf2009cyber}. The Internet was originally built as a best-effort service and thus not designed for real-time applications. Even though new solutions for real-time communications are always being worked out, it is problematic to be carried over the involved devices, e.g., routers. Therefore, instead of introducing changes at lower layers, a popular approach is to apply them at higher layers, especially the application layer. The transport protocol \textit{QUIC} is already actively used on the world wide web (WWW)~\cite{ruth2018first}. This protocol is designed to speed up the loading of Internet pages. However, due to its properties, QUIC can also be used for purposes outside the WWW~\cite{iyengar2021quic}. As more and more end devices have multiple network interfaces, e.g., Ethernet, Wi-Fi, and LTE, using them together for one connection generated a surge of interest in recent years. An obvious goal here is to bundle the available transmission rates of the interfaces in order to achieve higher throughput. For shorter transmissions, however, the path selection, i.e. whether a path should be used or not, can play a decisive role here. There are also approaches to transmit data redundantly, for example by sending duplicates. This is intended to reduce latency and can be particularly interesting for real-time applications because packet losses are compensated. However, it has been shown that sending all data redundantly can reduce latency at the transport layer, but not necessarily at the application layer. In addition, redundant transmissions are associated with higher data consumption. QUIC also offers the possibility of transmitting different data over a single connection. Different data may have different requirements, so it may be worthwhile to prioritize some of the data and transfer some of them redundantly. 

Time-critical applications that work together with sensors need low latencies and reliability. Some typical applications are autonomous driving and haptic communication. The QUIC transport protocol has promising approaches that can be used to reduce latencies for time-critical data. In particular, the multipath extensions that have been developed offer even more possibilities. This work primarily considers which measures can be used to transmit periodically generated data, such as that supplied by sensors, with the lowest possible latency via multipath QUIC. With the aforementioned possibility of transmitting data redundantly, the latency of such data can be kept low and data losses can be compensated leading to an increase in reliability. In addition to reliable transmission, however, there are other factors that must be taken into account. In the case of exchanging both time-critical and less-critical data, the priority of the important data with a low transmission delay needs to be guaranteed, as well as, a low latency between the generation time and transmission.

The remainder of this paper is organized as follows. Firstly, a background to some definitions and state-of-the-art is given in section~\ref{Section2}. The implementation and analytical model are described in section~\ref{Section3}. In section~\ref{Section4}, the evaluation results with different features and parameters are shown. Finally, section~\ref{Section5} concludes
the paper.

\section{Background}
\label{Section2}
The main transport protocols that can be used on the Internet are TCP and UDP~\cite{iren1999transport}. Establishing a new transport protocol is difficult because it needs to be available on the end devices as well as not be restricted by middleboxes~\cite{papastergiou2016ossifying}. In order to nevertheless develop a new transport protocol, but at the same time take the problem of ossification into account, Google introduced the \textit{QUIC} (Quick UDP Internet Connections) protocol in 2013. Although it is a transport protocol, it is actually implemented in the application layer and relies on UDP~\cite {Langley2017TheQT}. Since more and more devices have multiple network interfaces, the multipath TCP extension was proposed and implemented for TCP, which allows multiple interfaces and thus multiple paths to be used for one connection~\cite{paasch2014multipath}. In 2017, a multipath extension has also been presented for QUIC together with a prototype~\cite{de2017multipath}. In contrast to multipath TCP, Multipath QUIC (MP-QUIC) is subject to fewer limitations by middleboxes and is therefore more flexible.

To measure the performance in the network, there are different metrics like transmission delay and loss. The duration required for a packet to be sent from the sender to the recipient and back again is Round Trip Time (RTT). In QUIC, this time is determined by the time at which a packet was sent and the time at which the corresponding acknowledgment was received. The delay from the sender to the receiver (forward One-Way-Delay (OWD)) is not necessarily identical to the delay from the receiver to the sender (reverse OWD). In this work, however, for simplicity, it is assumed that the forward OWD is identical with the backward OWD, i.e. $OWD = \frac{RTT}{2}$.

The congestion control regulates the sending rate to avoid overloading the network. A receiver can use acknowledgments to confirm the successful delivery of a packet to the sender. A missing acknowledgment indicates that a packet has been lost. This mechanism is used by loss-based congestion control procedures. A lost packet is interpreted as an indication that the network is overloaded. The sender then throttles its transmission rate. The loss of a packet can also be caused by a transmission error, especially in wireless networks, and in this case, would not necessarily be an indication of overload. There are works, e.g., \cite{han2019machine}, in which attempts are made to determine the cause of losses and adjust the transmission rate accordingly. The Congestion Window (cwnd) is a component of congestion control, which specifies the number of unacknowledged packets or bytes that are allowed to be in circulation at the same time. Since the number of bytes or packets that can be in circulation is limited by the congestion control, there might be some data that can't be sent immediately. This means that the data will not be sent until the congestion window has a corresponding free space again. This causes transmission delay, which is a limitation for time-critical data, and motivated us to propose a new method to reduce the latency and loss for data with higher priorities or time-critical data.

\cite{vu2020latency} considered a method called selective redundancy. Here it is argued that duplicating as many packets as possible keeps latency at the transport layer low, but not necessarily at the application layer. The solution \textit{Selective Redundant MP-QUIC} was proposed first in~\cite{mogensen2019selective} as well as in~\cite{mogensen2018reliability}. They propose low latency and redundant transmission for high-priority data. They will always send data with priority regardless of the state of the congestion window. However, to avoid overriding the operation of the actual congestion control procedure, the congestion window is decreased by the appropriate number of bytes for the next interval and then reset to its previous value. Temporarily ignoring congestion control for data with priority is certainly acceptable for such small amounts of data ($1200\,$Byte). However, if larger amounts of data are to be transferred, for example, $20\,$kB, this approach could override the principles of congestion control. In this paper, we, therefore, implement an alternative solution that does not ignore the predefined limits of congestion control, but instead announces early that certain capacities in the congestion window must not be used by data without priority.

\section{The analytical model}
\label{Section3}
\begin{figure}[htb]
\centering
  \includegraphics[width=0.9\linewidth]{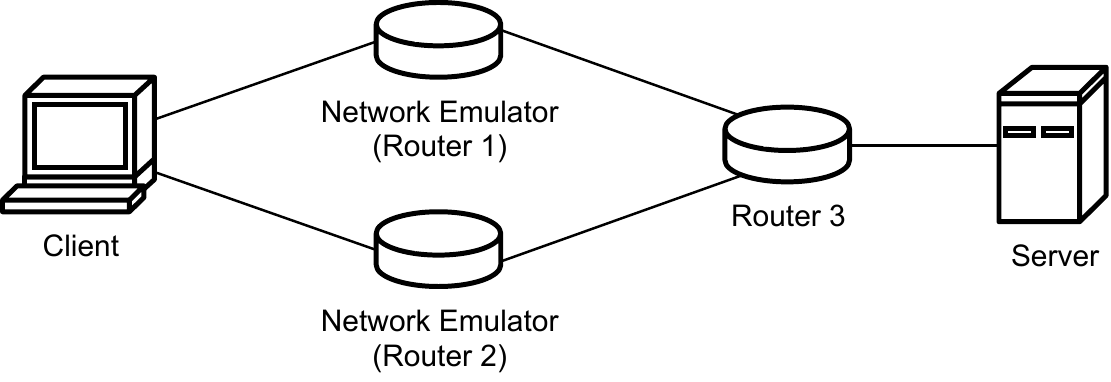}
  \caption{Used topology with two paths}
  \label{testbed:topologie}
\end{figure}

In this work, a traffic generator, stream selector, stream scheduler, and path scheduler are implemented. The topology shown in Figure~\ref{testbed:topologie} is provided for performing the experiments. The data transfers are performed in an emulab instance~\cite{emulab}.

\subsection{Traffic generator}
In order to emulate periodically sending data sources such as sensors, a client-server application is implemented. This makes it possible to define multiple data sources and to send and receive their generated messages. Here partial concepts from~\cite{vu2020latency} are adopted. Data sources can be defined in the traffic generator. These generate messages at equal time intervals, which are then written to streams. These streams are all transmitted over the same QUIC connection. Generally, we categorize the messages into priority data and background data. To generate priority data, data sources are set up in the traffic generator with a fixed inter-arrival time and message size. The priority data rely on low delays due to their importance. The background data have no special time requirements and therefore do not have to be considered individual messages, but the throughput is of interest here. The goal is that messages with priority experience low delays, while background data should simultaneously achieve high throughput by using two paths.

\subsection{Stream selector}
When a data source generates a message, the server application writes it to an appropriate stream. Which stream is used for this is decided by the stream selector. One possibility is to use a single stream for background data and another stream for priority data. The disadvantage of this approach is the occurrence of \textit{intra-stream head-of-line blocking}~\cite{marx2019resource}.

In this work, the target is that no intra-stream head-of-line blocking should occur between two messages. A packet loss for one message should not affect other messages. Therefore, only one message may be transmitted on a stream at the same time. This is realized by the fact that the complete reception of a message must always be acknowledged. When the client has completely received a message on a stream, it sends back a byte on the same stream. Once the server has received this byte, this stream can be used for another message. The signaling that a stream is free again is implemented here at the application layer.

\subsection{Stream scheduler}
When a data packet is created, it must be decided from which stream data will be taken. The stream scheduler is responsible for this assignment. Since QUIC packets are independent of their content, in principle data from several different streams can be contained in a single packet. However, the QUIC standard~\cite{ietf-quic-transport-04} advises against this, because the loss of a packet can then affect multiple streams, see also~\cite{marx2019resource}. Therefore, only data from one stream is transmitted in one packet here. A packet thus always contains only one stream frame at most. However, other frame types, for example, acknowledgment frames, can be contained in a packet in addition to a stream frame.

\begin{figure}[htb]
  \centering
  \subfigure[Round-Robin stream scheduler]{
    \includegraphics[width=1\linewidth]{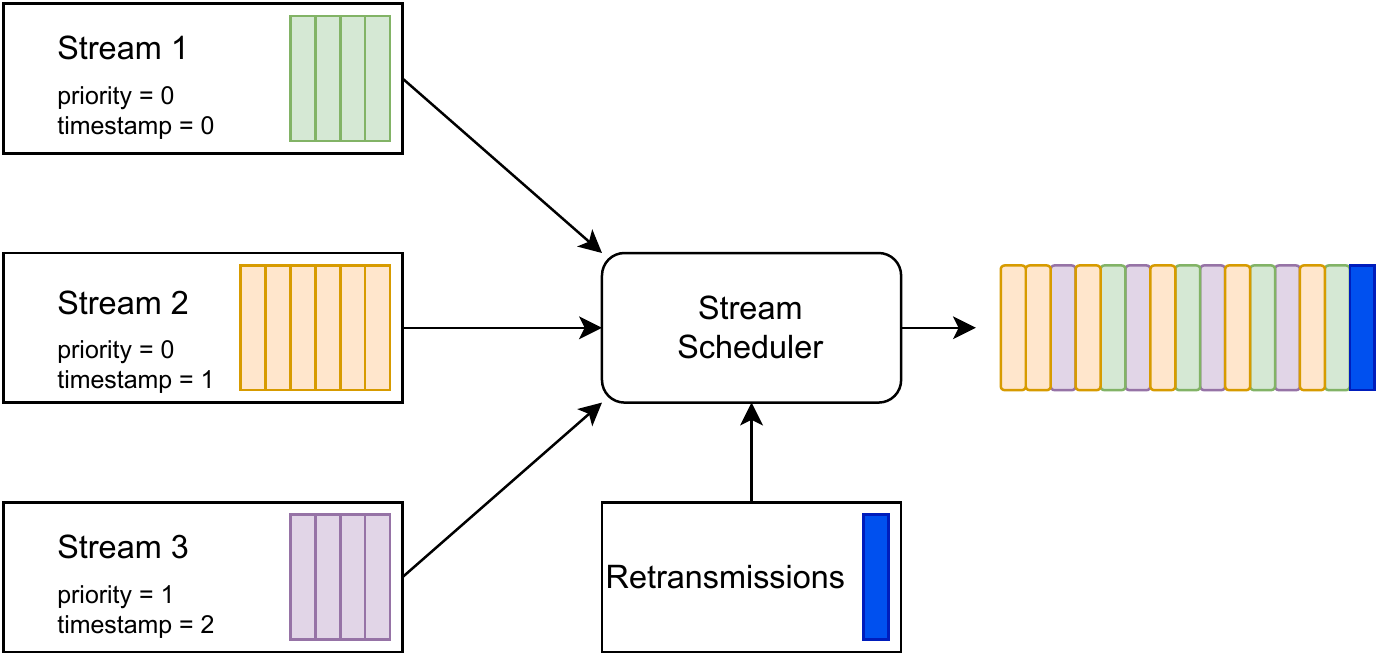} 
    \label{fig.2.1}
    }
    \subfigure[Priority FIFO stream scheduler]{
    \includegraphics[width=1\linewidth]{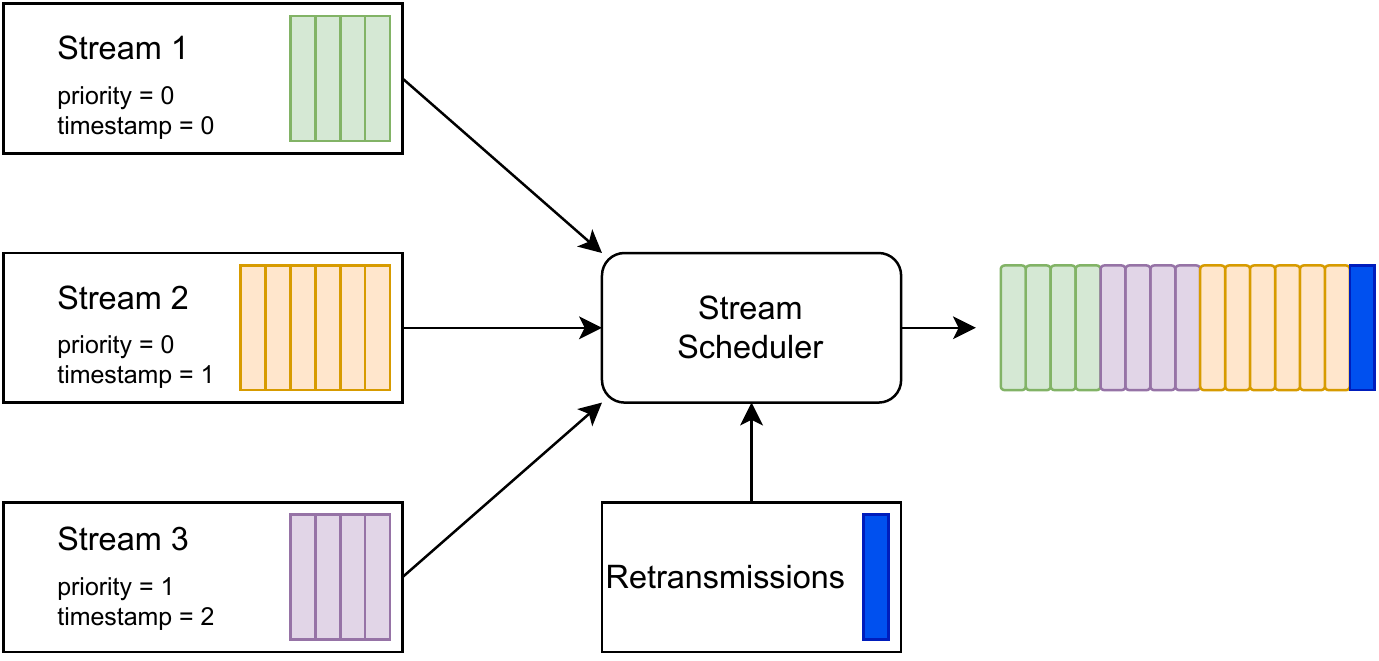} 
    \label{fig.2.2}
}
  \caption{Round-Robin vs. priority FIFO stream scheduler}\label{tg:gfx:vergleichStreamScheduler}
\end{figure}

A round-robin stream scheduler is already implemented in MP-QUIC. It treats each stream in the same way. In the present work, data is abstracted into messages. In addition, messages can have priority. Therefore, the priority FIFO stream scheduler is implemented. The default round-robin stream scheduler and the priority FIFO (First In - First Out) stream scheduler implemented here are compared in Figure~\ref{tg:gfx:vergleichStreamScheduler}. In the priority FIFO stream scheduler, retransmissions are always processed first, regardless of their priority. If no retransmissions are available, streams with a priority bit set are prioritized. If no messages with priority are available for transmission, the remaining streams are processed. Within these sets of streams, the FIFO principle is applied. 

If an excessive amount of priority data is generated, the strict prioritization can affect to streams without priority which can no longer be processed. This is also referred to as \textit{starvation}~\cite{jonglez2020end}. In the present work, however, the size of the priority data is always chosen so that this problem can be neglected.

\subsection{Priority path scheduler}
We present a new path scheduler that avoids the transmission delay for periodically generated priority data.  A \textit{Congestion Window Reservation} is intended to ensure that there is sufficient space in the congestion window at the time priority data is generated. To avoid delays due to packet loss, a redundancy procedure is also used that duplicates priority data and sends it on both paths. 

\subsubsection{Congestion window reservation}
To avoid transmission delay for priority data, the Congestion Window Reservation (CWR) scheduler is implemented. This uses a reservation procedure that keeps space in the congestion window for priority data free. It may be worthwhile to wait for priority data to be generated instead of sending background data that excessively occupy the available congestion window. This is to ensure that messages with priority find sufficient free congestion immediately after their generation. For data without priority, it follows that an actually free area in the congestion window may remain unused. This results in periods during which no data are sent, although this would be possible from the point of view of congestion control. The reservation procedure is based on the assumption that the congestion window remains constant. Neither growth nor reduction is taken into account. If a packet loss occurs before a message is generated, the reserved area cannot be used because it is dropped.

The CWR scheduler therefore always assigns background data to the path with the lowest RTT on which no reservations are at risk. For priority data, the path with the lowest RTT and sufficient space in the congestion window is selected.

\begin{figure}[htb]
\centering
  \includegraphics[width=1\linewidth]{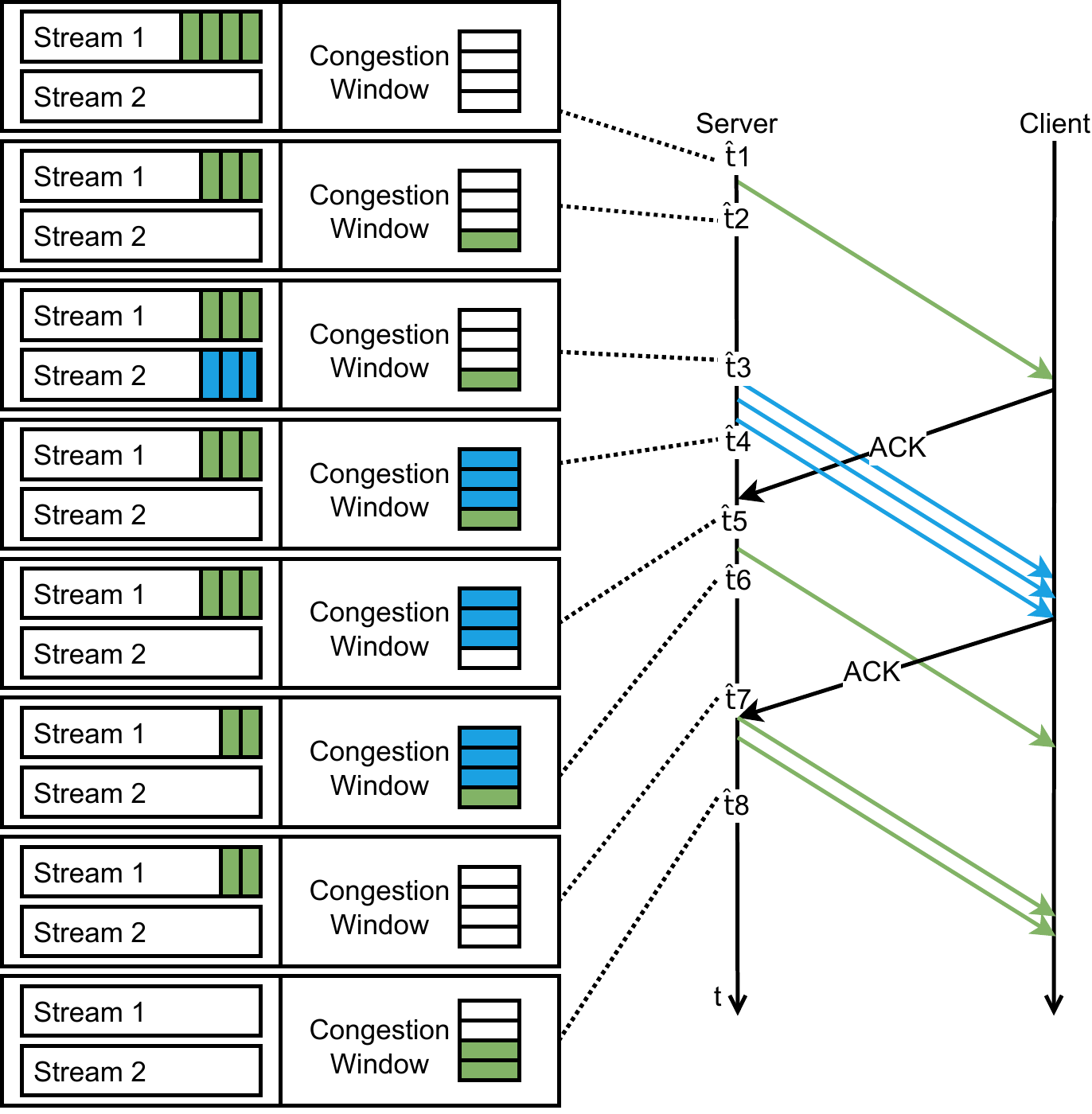}
  \caption{An example of the scheduler}
  \label{tg:gfx:senddelay2}
\end{figure}

The scheduler proposed here is illustrated in an example in Figure~\ref{tg:gfx:senddelay2}. Here, the congestion window has space for $4\,$ packets and is empty at the beginning. At $t_1$, data are available in stream~1 for sending. It is assumed that at time $t_3$ data with priority are written to stream~2, whose size corresponds to $3$ packets. Therefore, only one packet is generated from the data of stream~1 and sent at $t_2$, leaving space for three packets in the congestion window. At the time $t_3$, the data in Stream~2 are then ready to be sent. Since the congestion window is kept free up to this point, the sending happens immediately and is completed shortly after $t_4$. The remaining data from Stream~1 is sent at the next possible times $t_5$ to $t_8$.

\subsubsection{CWR + Redundancy}
The option to transmit data redundantly with priority is now to be made available. The CWR+RED (Congestion Window Reservation + Redundancy) is implemented for this purpose. While the CWR scheduler adds a reservation to only one path, the CWR+RED adds the reservations to both paths. To send data with the priority, the corresponding packets are duplicated and sent on all paths. Due to the reservation on all paths, there is no transmission delay even for the duplicates. This scheduler behaves exactly like the CWR scheduler on the background data.

In case the reservations are not successful and the congestion of each path is too small to send a message completely, the scheduler will refrain from redundant sending. There is then still the possibility that the sum of the free areas in the congestion windows is sufficient to send the data. In general, for packets that could not be duplicated immediately, no duplicate is created afterward.  

\section{Results}
\label{Section4}
In this section, the evaluation results of CWR and the CWR+RED scheduler will be provided. It will be examined whether delays for periodically generated priority data can be reduced by using the reservation and redundancy method, whereas the influence on the background data will be considered. For the experiments, the path schedulers LowRTT, CWR, and CWR+RED are compared, while the stream scheduler Priority FIFO is used in all cases.

\subsection{Used metrics}
The used network parameters in the experiments are RTT, transmission rate, and loss rate. The RTT is chosen depending on the different experiments, while the transmission rate and loss are set to $100\,$Mbit/s, and $0,05\,\%$, respectively. An important assumption made here is that the congestion windows of the paths have a sufficient size at all times, which in principle allows enough space to be kept free for priority data.

The Message Completion Time (MCT) and the throughput are the main metrics used in the evaluation. MCT indicates the time from the generation of a message at the server to its complete reception at the client application. This time includes all delays that can occur during the transmission of a message. For the representation of the MCTs, the Complementary Cumulative Distribution Function (CCDF) is used. The throughput indicates how much data has been transferred at the transport layer per unit of time. This includes both priority and background data.

\subsection{One priority data source}
\label{tg:evaluation:einedatenquelle}
\begin{figure}[htb]
\centering
\subfigure{
 \includegraphics[width=1\linewidth]{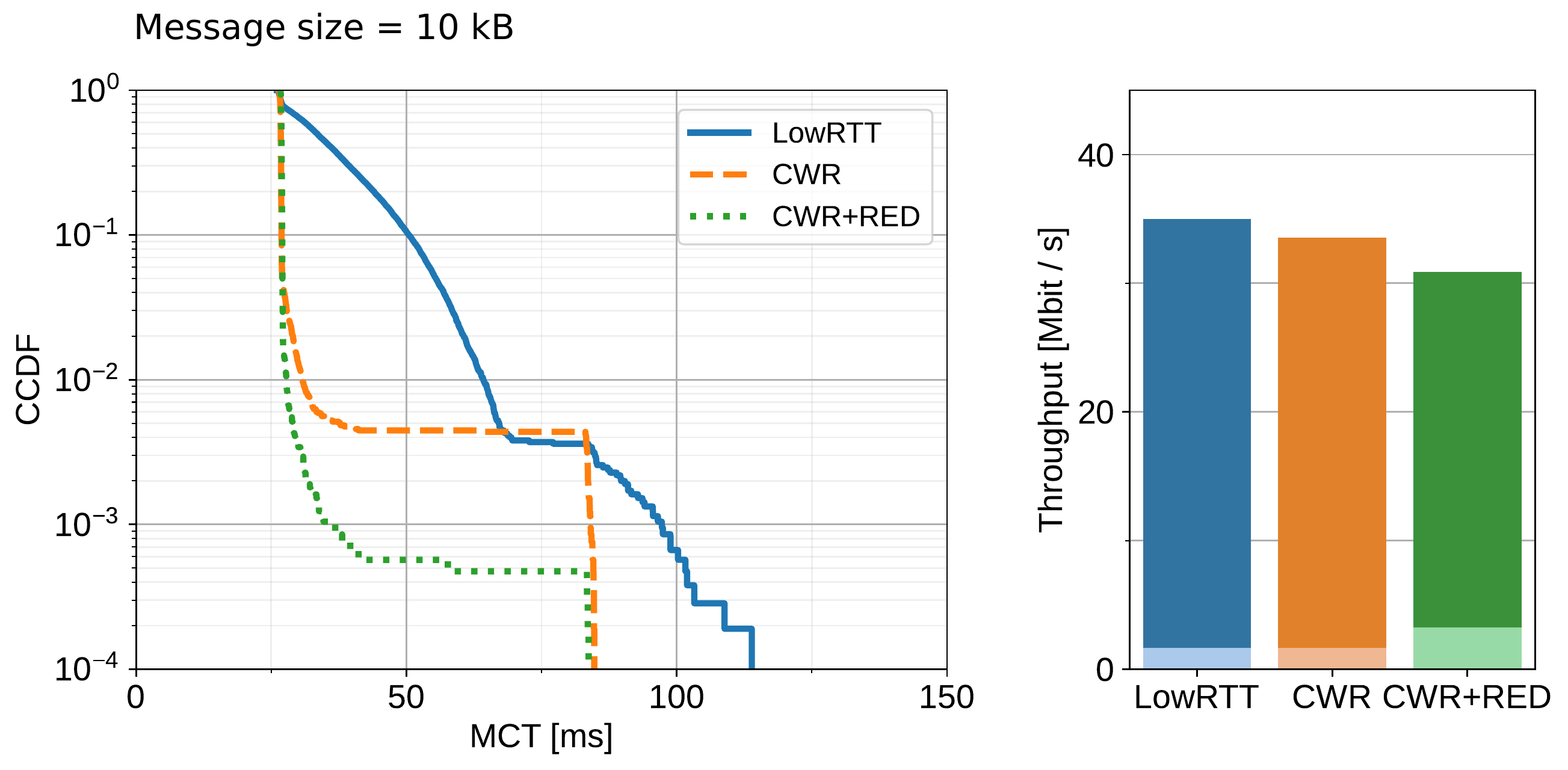}
}
\subfigure{
 \includegraphics[width=1\linewidth]{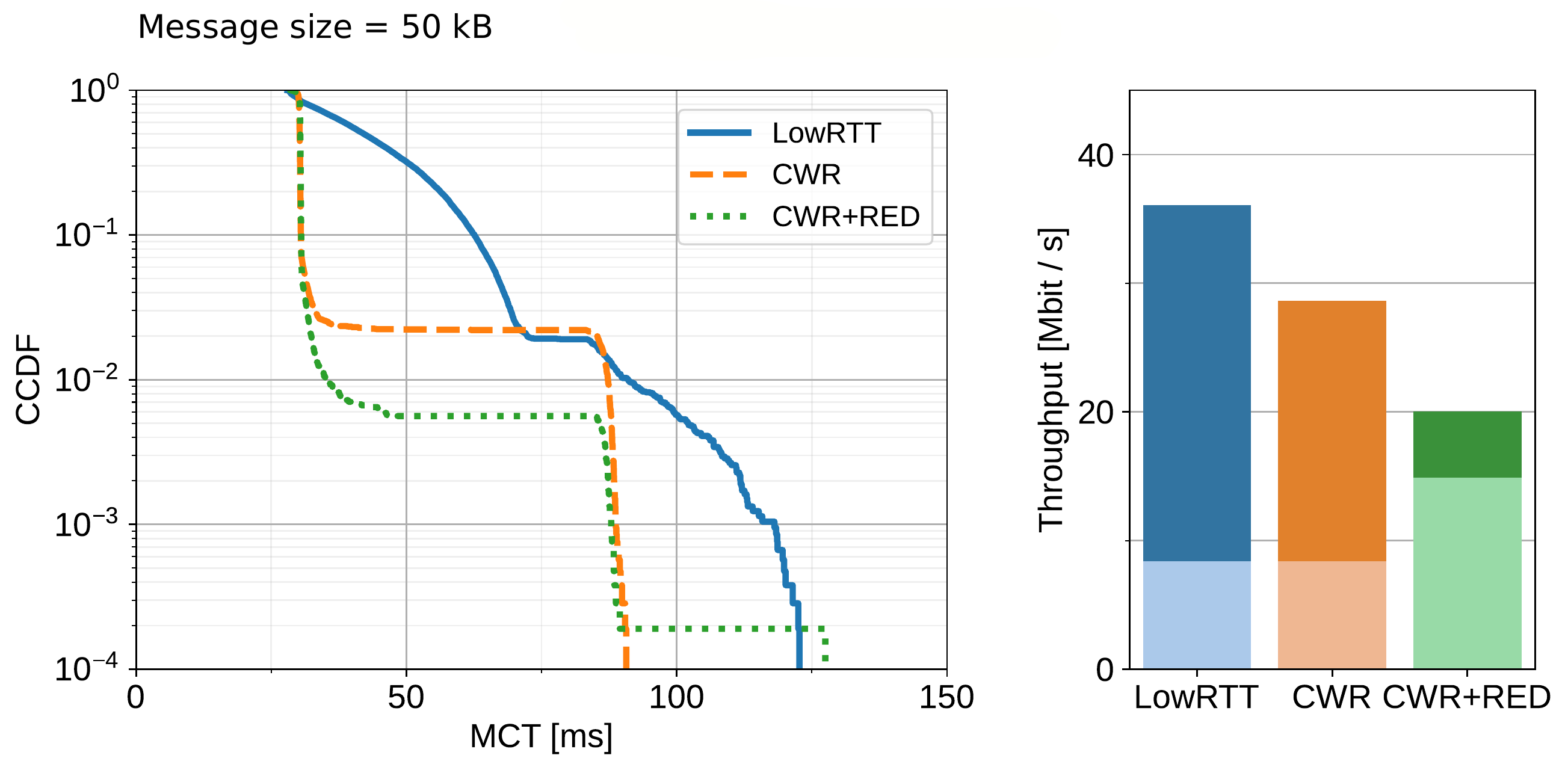}
}
\caption{Distribution of MCTs of messages with priority (left) and throughput of the whole transmission (right).}
\label{tg:eval:throughput1}
\end{figure}

We evaluate the mentioned schedulers in different experiments while sending priority messages only from one data source. The results are illustrated in Figure~\ref{tg:eval:throughput1}. On the left, the CCDF of the MCTs for the priority data is shown in each case. On the right, the throughput is represented together with the share of the priority data. Lighter areas indicate the share of priority data in the total throughput.

It can be seen that the use of LowRTT schedular can result in larger MCTs for priority data. This scheduler treats priority and background data the same, so there may be transmission delays. In comparison, the CWR scheduler (without redundancy) shows significantly lower MCTs, since the priority data are usually sent immediately after they are generated. The MCTs in this case are about 1 OWD which corresponds approximately to OWD = RTT/2 = 25 ms. It can also be seen that some MCTs due to packet loss are over 1 RTT. In the case of packet loss, the Early Retransmision method will resend the packet, which takes at least 1.625 RTT or 81.25 ms.

The CWR+RED scheduler, which duplicates priority packets on both paths, shows low MCTs for most messages. In rare cases like packet loss of a packet and its duplicate, there are higher delays. Another reason would be that in some transmissions the reservation is not successful. This happens if a packet loss occurs shortly before the generation of a message which is due to the fact that the congestion window has got reduced. In that case, the reserved area may be dropped.

As the message size increases, packet loss is more likely, which describes why slightly more MCTs lie far above the OWD. All in all, the reservation procedure works even if the message size increases. However, reductions in throughput occur when the CWR and the CWR+RED scheduler are being used. It can be seen that the share of the priority messages is much higher when the CWR+RED scheduler is used. The reason is, that this scheduler duplicates as much priority data as possible. With increasing the message size, no change can be seen for LowRTT, since the priority data, regardless of their size, do not affect the operation of the scheduler.

It has been shown that using the CWR and CWR+RED scheduler, the MCTs can be kept low, but the overall throughput is reduced. Since the throughput using the LowRTT scheduler does not change while changing the message size or inter-arrival time, the cause must lie in the reservation procedure. This reserves in the congestion window exactly as many bytes as are needed for the messages, but does not explicitly modify the size of the congestion window. Therefore, it is reasonable to look at the growth of the congestion window in the performed experiments. The average increase in congestion windows per RTT (50 ms) during the congestion avoidance phase is calculated. The results are shown in Table~\ref{tg:tabelle:wachstum}.

\newcolumntype{P}[1]{>{\centering\arraybackslash}m{#1}}
\begin{table}[ht!]
\centering
\scalebox{0.9} {
\begin{tabular}{|P{1.1cm}|P{1.7cm}|P{1.8cm}|P{1.6cm}|P{1.6cm}|}
\hline 
    Inter-Arrival time (ms) & Message size (kB) & Scheduler &  $\Delta cwnd_{1,CA}$ (Byte/ RTT) & $\Delta cwnd_{2,CA}$ (Byte/ RTT)  \\
    \hline{}
     & & LowRTT & $1342$ & $1340$ \\   \cline{3-5}
    $50$ & $10$  & CWR & $1264$ & $1263$ \\   \cline{3-5}
     & & CWR+RED & $1190$ & $1189$ \\ 
    \hline{}
     &  & LowRTT & $1333$ & $1334$ \\   \cline{3-5}
    $50$ & $25$ & CWR & $1175$ & $1246$  \\   \cline{3-5}
    & & CWR+RED & $969$ & $972$\\
    \hline{}
    & & LowRTT & $1336$ & $1350$  \\   \cline{3-5}
    $50$ & $50$ & CWR & $1111$ & $1102$ \\   \cline{3-5}
    & & CWR+RED &  $738$ & $771$ \\
    \hline{}
    & & LowRTT & $1346$ & $1354$ \\   \cline{3-5}
    $100$ & $10$ & CWR & $1294$ & $1318$ \\   \cline{3-5}
    & & CWR+RED & $1289$ & $1306$\\
    \hline{}
     & & LowRTT & $1334$ & $1345$ \\   \cline{3-5}
    $100$ & $25$ & CWR & $1288$ & $1254$ \\   \cline{3-5}
    & &  CWR+RED & $1183$ & $1200$ \\
    \hline{}
     & & LowRTT & $1332$ & $1335$ \\   \cline{3-5}
    $100$ & $50$ & CWR & $1149$ & $1183$ \\   \cline{3-5}
    & & CWR+RED & $1048$ & $1033$ \\
    \hline
\end{tabular}
}
\caption{Mean growth of congestion windows per RTT}
\label{tg:tabelle:wachstum}
\end{table}
In fact, the growth of the congestion window is limited when using the reservation procedure. In the congestion avoidance phase, the congestion window is increased by a maximum packet length when the amount of the acknowledged data is equal to the current size of the congestion window. The maximum packet length here corresponds to $1350\,$byte. Since an ACK (Acknowledgment) arrives about $1\,$RTT after a packet has been sent, the congestion window of a path therefore ideally grows by $1350\,$Byte per RTT at a full load. To achieve this, the space freed by an ACK must be reused immediately. This is the case with the LowRTT scheduler, which achieves almost the ideal gradient. However, the reservation procedure intentionally prevents the use of the free area in the congestion window and waits for the priority data. This may also increase the time required to send an amount of data equal to the size of the congestion window. Therefore, the table also shows a reduced growth for the CWR scheduler. The CWR always appends a reservation only to the path with the lower RTT. Because the paths have the same RTT, the choice of the path changes. Accordingly, growth is restricted on both paths. The CWR+RED scheduler always uses both paths for a reservation, which is why growth is restricted even more.

\subsection{Multiple priority data sources}
The aim now is to evaluate whether the method can also achieve a low transmission delay for priority messages from multiple data sources. For this purpose, three data sources with different inter-arrival times and message sizes are used simultaneously. The characteristics are given in Table~\ref{tg:evalNeu:tabelleX2}. 

\begin{table}[h]
    \centering
\scalebox{1}{
    \begin{tabular}{c|c|c}
    Data source  & Inter-Arrival Time & Message size \\
    \hline  
    $S_1$ &  $100\,$ms & $10\,$kB \\ 
    $S_2$ &  $70\,$ms & $7\,$kB  \\
    $S_3$ &  $135\,$ms & $5\,$kB 
\end{tabular}
}
    \caption{Parameters of priority data sources $S_1$, $S_2$,and $S_3$.}
    \label{tg:evalNeu:tabelleX2}
\end{table}

\subsubsection{Paths with the same RTT}
\begin{figure}[htb] 
\centering
\subfigure{
\includegraphics[width=1\linewidth]{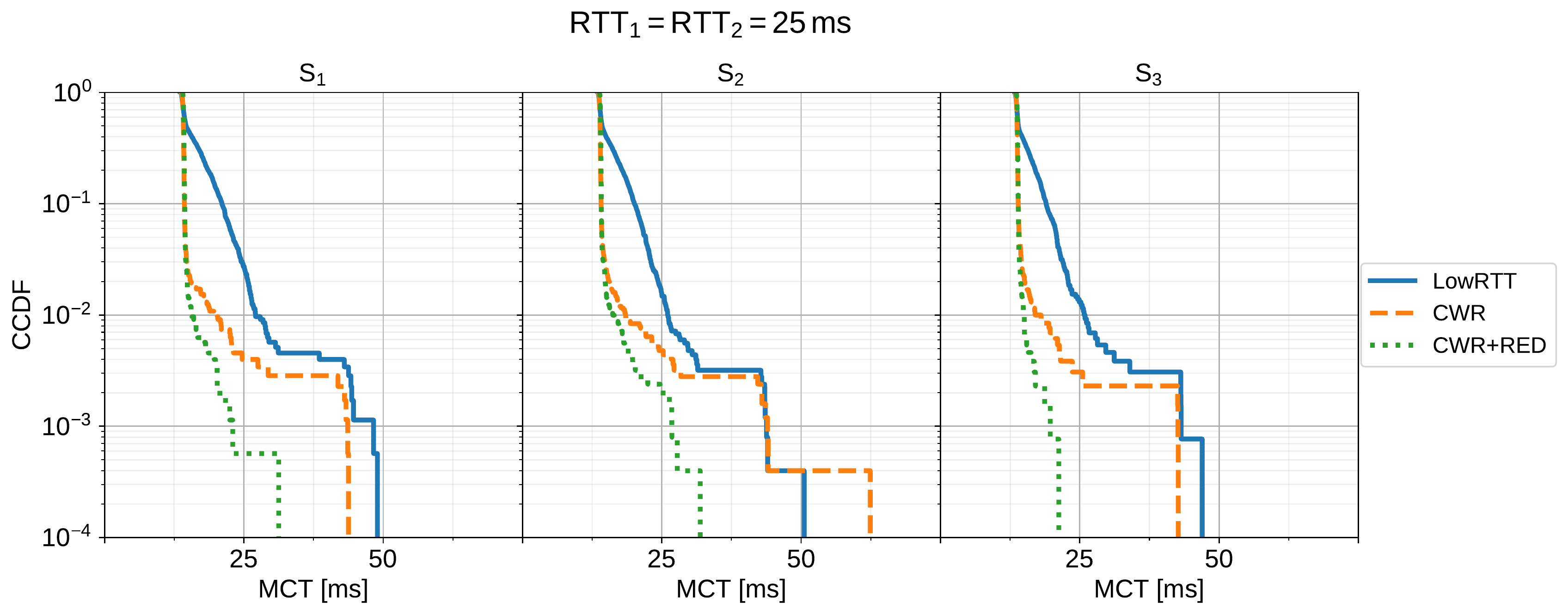}
}
\subfigure{
 \includegraphics[width=1\linewidth]{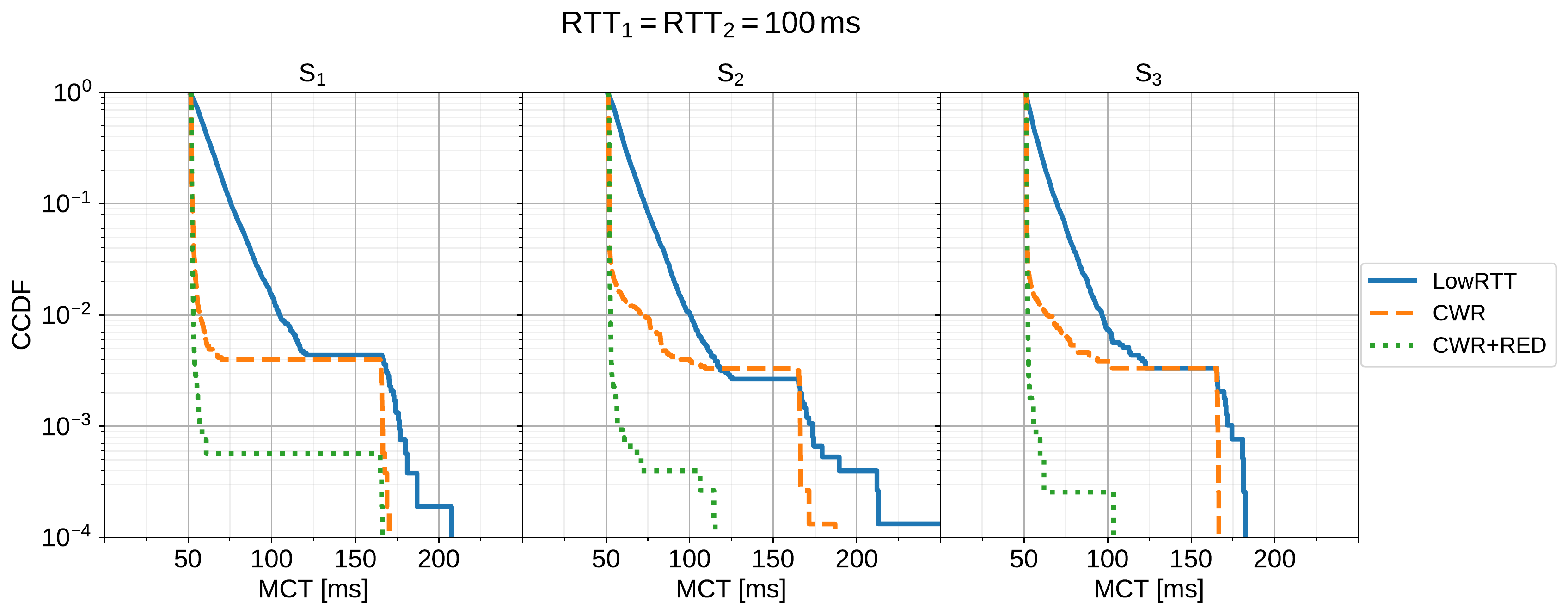}
}
\caption{Distribution of MCTs with equal RTTs for paths}
\label{tg:eval:mehrereDatenquellen}
\end{figure}

First, the experiments are performed with the same RTT on both paths. The MCTs for each data source are shown in Figure~\ref{tg:eval:mehrereDatenquellen}. 

The results show that the reservation method can also be used for multiple data sources with priority and reduces the transmission delay. For each individual data source, low MCTs are shown when the CWR scheduler is used. These are in the range of $1\,$OWD. The CWR+RED scheduler avoids the consequences of packet loss and provides high reliability and thus helping to keep an even larger fraction of MCTs low.

\subsubsection{Paths with different RTT}
\begin{figure}[htb]
\centering
\subfigure{
 \includegraphics[width=1\linewidth]{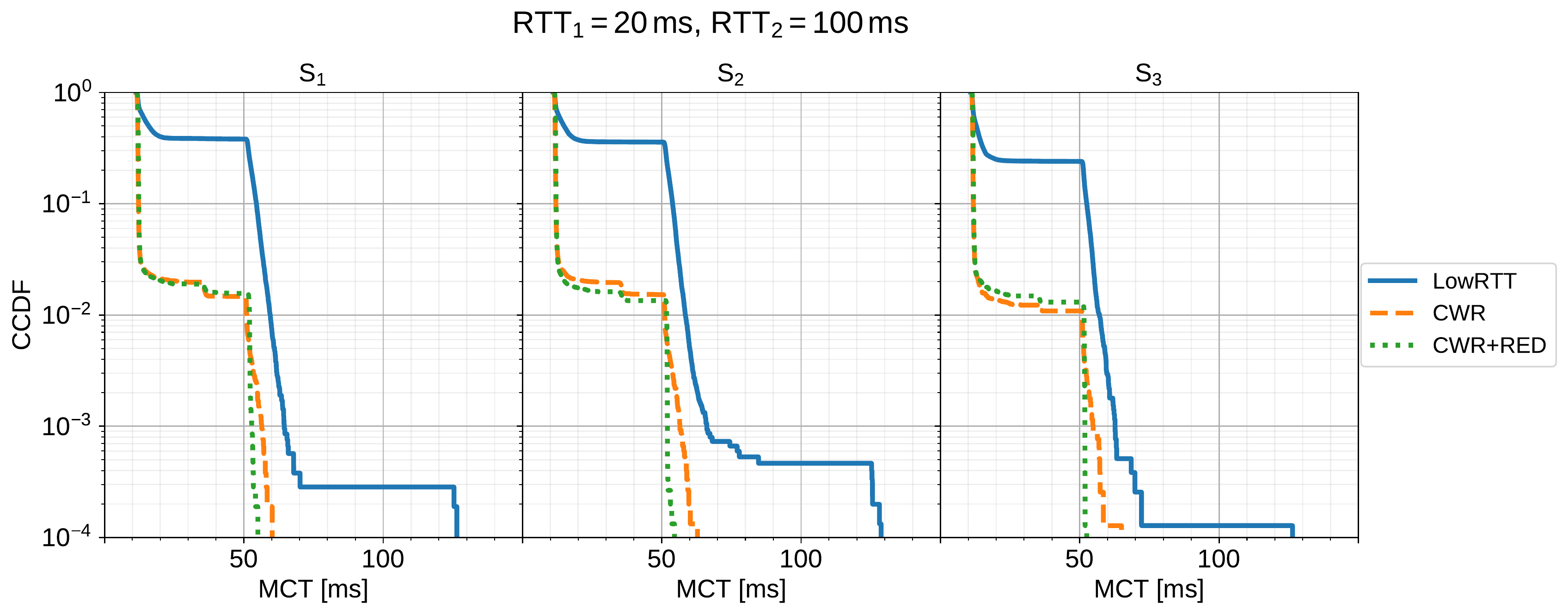}
}
\subfigure{
 \includegraphics[width=1\linewidth]{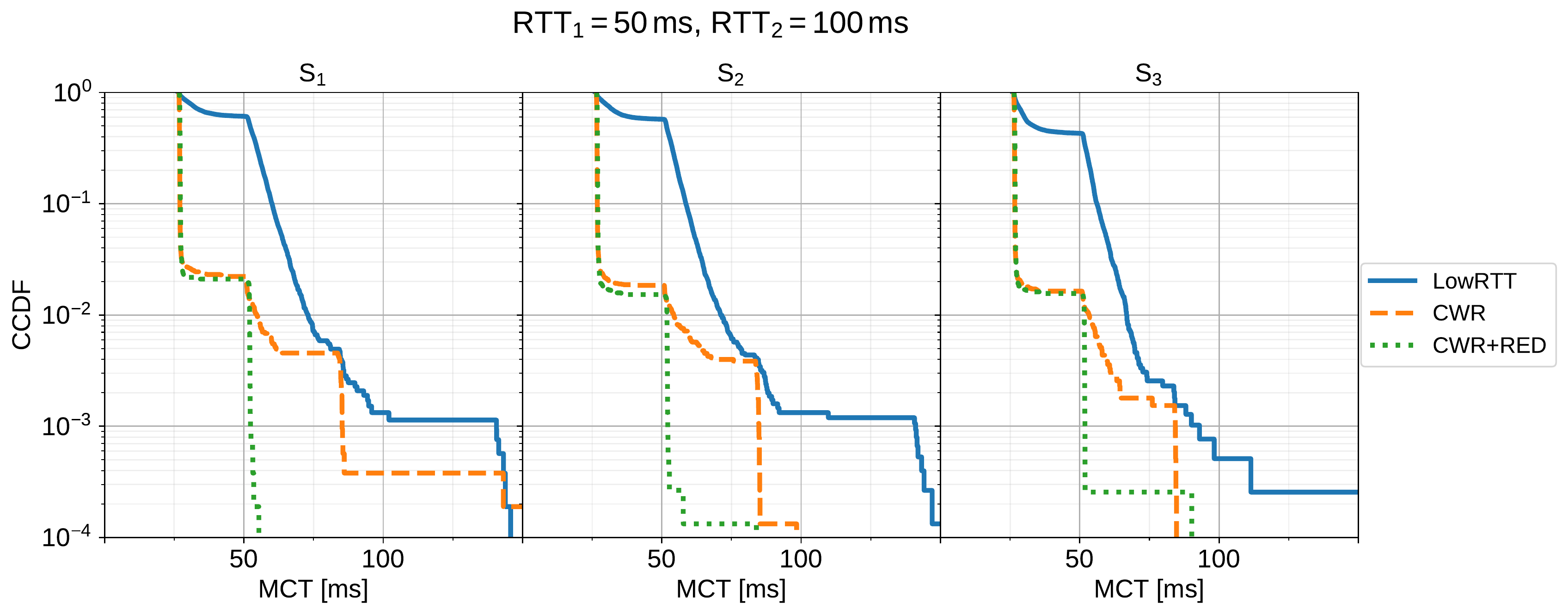}
}
\caption{Distribution of MCTs with different RTTs for paths}
\label{tg:eval:iat4:inhomogen}
\end{figure}

Now we consider paths with very different RTTs. The transfers are performed once at $\mathrm{RTT_1}=50\,$ms and $\mathrm{RTT_2}=100\,$ms, and then at $\mathrm{RTT_1}=20\,$ms and $\mathrm{RTT_2}=100\,$ms shown in Figure~\ref{tg:eval:iat4:inhomogen}. 

Since the CWR scheduler always performs reservations on the path with the lower RTT, it is immediately available for the messages with priority if the reservation is successful. The MCT is then, except in the case of packet loss, about $\mathrm{OWD}_1$, i.e., the one-way delay of the first path. If the reservation is not successful, the scheduler operates like LowRTT. It may then happen that the slower path is selected and the MCT will be $\mathrm{OWD}_2$. In case of packet loss, the time will increase. Using the CWR+RED scheduler in case of packet loss on the first path, the redundancy procedure can recover the loss via the duplicate on the second path. The CWR+RED scheduler therefore still shows advantages in some cases. For $\mathrm{RTT_1}=20\,$ms and $\mathrm{RTT_2}=100\,$ms, the distribution of MCTs is nearly identical for CWR and CWR+RED. The RTTs are so far apart that the redundancy procedure is not worthwhile at this point. A packet loss on path~1 is resolved with the Early Retransmision procedure even before the duplicate arrives on path~2. Depending on how large the difference between the RTTs is, duplicating packets may make more or less sense.

\section{Conclusion}
\label{Section5}
In this work, we investigated the possibilities of achieving low latency for time-critical data in multipath QUIC. We considered the case where priority data and background data are to be transmitted on the same multipath QUIC link. To reduce the transmission delay and compensate for packet loss, a reservation procedure and a redundancy procedure have been implemented as well. In addition, a stream scheduler was implemented that always prioritizes time-critical data. The main purpose of this is to ensure that the reserved area in the congestion window is only used by data with higher priority. The evaluation has shown that the combination of reservation and redundancy procedures can keep message completion times very low so that they are in the range of the one-way delay and exhibited higher reliability for time-critical data as well.

\bibliographystyle{IEEEtran}
\bibliography{references}

\end{document}